# Comparative Analysis of Two Astrometric Measuring Methods on Five Known Binaries

Syfrett, Malachi [1], Major, John [2] Gamage, Gihan [1]

[1] New Mexico State University - Alamogordo, New Mexico; (msyfrett@nmsu.edu)
[2] Colorado Mountain College, Steamboat Springs, Colorado

## Abstract

Data on five Washington Double Star catalog binaries were collected from the Dimension Point Observatory (Mayhill, New Mexico) and Las Cumbres Observatory (Cerro Tololo, Chile) on February 19, 2025, and March 5, 2025, respectively. Student researchers participating in the Four Corners Research Seminar measured the position angle (θ/deg) and Separation (ρ/arcsec) of each target using AstroImageJ and an author-created script utilizing the Astropy module in Python3. Each target was imaged using a variety of instruments, filters, and exposure times. Compared to the extrapolated 6th Orbit Catalog estimates, measurements using AstroImageJ were within 1.52% of θ and 14.87% of ρ, while the author's automated code method provided measurements within 4.09% of θ and 16.59% of ρ.

## 1. Introduction

In January 2025, students from four community colleges and one high school joined to form the Four Corners Research Seminar organized by Paul McCudden of Colorado Mountain College to observe and measure binary stars. These teams were organized with instructors and members of the professional astronomy community from the states of Arizona, Colorado, New Mexico, and Utah. Supported by a grant from the National Science Foundation (NSF#2428684), these teams worked with observatories in the region and further afield to gather data on known binary systems.

Remote access to small-aperture, near professional-grade telescopes, coupled with free tools like AstroImageJ, Python3, and Google Workspace, has made it possible for students and amateur astronomers to contribute scientifically valuable binary star observations (Faughn et al. 2023). From this emerged two methods for analyzing the images: a manual method and a semi-automated method. These two methods were used to measure position angle (θ) in degrees (°), separation between the primary and secondary stars (ρ) in arcseconds ("), and differential magnitude (ΔMag) values for five targets. These are hallmark measurements that help us understand the systems and can be used in many cases, such as determining orbital parameters. These values were compared between the two methods used in this study and against the predicted values from the United States Naval Observatory (USNO). Other qualities of the astrometric tools were considered, including reliability, ease of use, and efficiency in reductions.

### 1.1 Target Selection

The targets were selected from a list generated by Paul McCudden that combined the USNO Washington Double Star Catalog (WDS) and the Sixth Catalog of Orbits of Visual Binary Stars (6th Orbit) (Harkopf, 2001). The initial target list consisted of systems that were within Right Ascension (RA) 8.0hrs - 12.0hrs, Declination (DEC) +10° - +60°, with a predicted current Sep (ρ): 3.0" - 5.0", at least 20 total observations, and no observations within the last three years. Additionally, a ΔMag constraint was considered such that no ΔMag greater than three was included.

Then the initial list was filtered according to the specifications of the instruments at our disposal and the parameters outlined above, which reduced the list to five stars. Four targets were observed at Dimension Point Observatory and one at Las Cumbres Observatory Global Telescope (LCOGT). The LCOGT location allowed the addition of a target below the DEC +10 parameter mentioned above (WDS 08153-6255). The instruments and target list combinations led to an agreement of three main ideas: the team aimed to test the limits of the telescopes and cameras available based on resolution of systems with high differential magnitudes and/or low separations, provide information about stars that have not recently been observed, and determine whether one method of astrometric measurement was superior to the other.

The targets exhibit a wide range of characteristics (Tables 1 and 2). Distances vary considerably, ranging from approximately 40 light-years (ly) to about 260 ly, with orbital grades spanning from 3 to 5 (Table 1). Observations have been conducted as early as 1821 up to 2022, with the total number of observations ranging from 323 to 26. The periods range from 804 years to 3000 years, and the spectral classes vary from A2 to M1. All predicted separations for J2025 fall between 2" and 5" (except WDS 10067+1754), and no ΔMag values exceed 3.

WDS 10067+1754 was selected for its challenging characteristics with the intent of understanding the limitations of instruments and methods used in this study. WDS 10067+1754 was last observed in 2021 and features a significantly smaller separation of 0.435 arcseconds (Stelle Doppie). The system still has very few observations compared to the other four targets. The authors hoped to contribute useful measurements to the record, given the target's 27 observations from 1991 to 2021.

*Table 1: Stelle Doppie specification of the binary systems used in this study (Stelle Doppie).*

| WDS ID | System | First Obs. | Last Obs. | Total # of Obs. | Orbital Period (yrs) | Distance (ly) | Spectral Class | Orbital Grades |
|---|---|---|---|---|---|---|---|---|
| 08095+3213 | STF 1187 AB | 1829 | 2020 | 323 | 1385 | 210.99 | F2 | 5 |
| 08508+3504 | STF 1282 AB | 1821 | 2019 | 180 | 804 | 168.32 | F8 | 4 |
| 08153-6255 | RMK 8 AB | 1851 | 2016 | 26 | 3000 | 252.87 | A2V | 5 |
| 10067+1754 | HDS1457 AB | 1991 | 2021 | 27 | 2023 | 322.32 | G1IV | 5 |
| 08554+7048 | STF 1280 AB | 1823 | 2022 | 261 | 689 | 37.79[1] | M1V | 3 |

---

[1] Data was taken from Gaia Data Release 3 (DR3)

*Table 2: Coordinates and historical data of the targets. (Stelle Doppie) Mag 1 and Mag 2 are the magnitudes of the primary and the secondary star.*

| WDS ID | RA J2000 | Dec J2000 | ρ/" | θ/° | Mag 1 | Mag 2 | ΔMag |
|---|---|---|---|---|---|---|---|
| 08095+3213 | 08h 09m 30.45s | +32° 13' 18.8" | 3.07 | 20.8 | 7.19 | 7.98 | 0.79 |
| 08508+3504 | 08h 50m 44.28s | +35° 04' 15.4" | 3.459 | 278.8 | 7.59 | 7.76 | 0.17 |
| 08153-6255 | 08h 15m 15.91s | -62° 54' 56.2" | 4.073 | 69.8 | 5.27 | 7.62 | 2.35 |
| 10067+1754 | 10h 06m 40.82s | +17° 53' 42.6" | 0.435 | 45.2 | 8.29 | 10.95 | 2.66 |
| 08554+7048 | 08h 55m 24.82s | +70° 47' 39.2" | 3.878 | 359.1 | 8.81 | 9.10 | 0.29 |

## 2. Equipment and Observation Strategy

The primary source of data acquisition was from the Dimension Point Observatory (Observatory V42) in Mayhill, NM, at an elevation of 2164m (7100ft). Observatory V42 features a Planewave CDK24 f/6.5 telescope with a Kepler KL400 BI Complementary Metal-Oxide-Semiconductor (CMOS) camera, accessed through the expertise of Geoff Stone.

The research initiative also utilized the Las Cumbres Observatory Global Telescope (LCOGT) facility (Observatory W89), specifically the Aqawan A edifice, situated at an elevation of 2195m (7,200ft) in Cerro Tololo, Chile. LCOGT featured a DeltaRho 350 f/3 Optical Tube Assembly telescope with a QHY600PH Series CMOS camera as part of the project funded by the NSF (see the Introduction).

The exposure times were chosen between 0.5 seconds to 2.0 seconds according to the magnitude of the target, with the intent of not saturating the target object. The target 08153-6255, observed using LCOGT, took several attempts at 1.0 seconds and 0.75 seconds before usable data were gathered at 0.5 seconds.

Upon finalizing the targets, our observations were optimized with appropriate photometric filters. All targets observed by V42 were classified within spectral types ranging from F-type (white/yellow) to M-type (red). Hence, the Johnson-Cousins V filter was selected for observations. Due to the color and spectral class of the target 08153-6255, it was first observed using the Johnson-Cousins V (JC-V) filter. However, due to the high luminosity of the primary, we opted for the SLOAN i′ filter to attenuate the intensity of the primary star and highlight the secondary star.

## 3. Methods

The first astrometric method utilized the software AstroImageJ (AIJ). This software offers the ability to easily plate solve and visually find the stars. If the images were of good quality, AIJ was able to automatically determine the center of the stars, find ρ, θ, and ΔMag. However, many images lacked the requisite quality for this automation due to not being fully resolved. Hence, the center of the stars had to be found manually, along with manually calculating ρ and θ. This is accomplished by clicking on the visually-identified center of the primary star, then dragging the cursor to the center of the secondary. AIJ provides the measurements for ρ and θ on the display (See Figure 7).

*Table 3: Summary of observing equipment and number of images requested for this study.*

| WDS ID | Observatory Code | Filter | Exposure Time | Number of Images |
|---|---|---|---|---|
| 08095+3213 | V42 | V | 1.0/1.5 | 25/25 |
| 08508+3504 | V42 | V | 1.0/1.5 | 25/25 |
| 08153-6255 | W89 | V/i′ | 0.5 | 50/49 |
| 10067+1754 | V42 | V | 1.5/2.0 | 25/25 |
| 08554+7048 | V42 | V/I | 2.0 | 25/25 |

The second method was designed by author Malachi Syfrett, named "Photometric Method for Double Star Analysis" or PMDSA (See Appendix), using Python3 (v3.12 and v3.13) and the Astropy (Astropy Collaboration, 2022) module. Although this method still requires manual plate solutions for each image, it can automatically isolate the stars and find ρ, θ, and ΔMag values.

Both methods were unable to determine ΔMag values for the targets WDS 08508+3504 and WDS 10067+1754. Additionally, AIJ was unable to provide ΔMag values for the targets WDS 08095+3213 and WDS 08153-6255 (See Discussion).

### 4. Data

For each target, the median, mean, standard deviation (STDEV), and standard error of the mean (SEM) for ρ, PA, and ΔMag from each method are displayed in Tables 4 - 8, along with images of the targets, USNO 6th Orbit plot, and surface intensity plots created using PMDSA. A summary of median ρ, PA, and ΔMag values for all targets can be found in Table 9.

*Table 4: Measurements of 08095+3213 using the JC-V filter.*

|  | ρ/″ AIJ | ρ/″ PMDSA | θ/° AIJ | θ/° PMDSA | ΔMag PMDSA |
|---|---|---|---|---|---|
| MEDIAN | 2.10 | 1.94 | 19.13 | 21.76 | 0.41 |
| MEAN | 2.05 | 1.97 | 19.53 | 21.35 | 0.40 |
| STDEV | 0.25 | 0.60 | 3.55 | 6.95 | 0.13 |
| SEM | 0.04 | 0.09 | 0.51 | 1.07 | 0.02 |

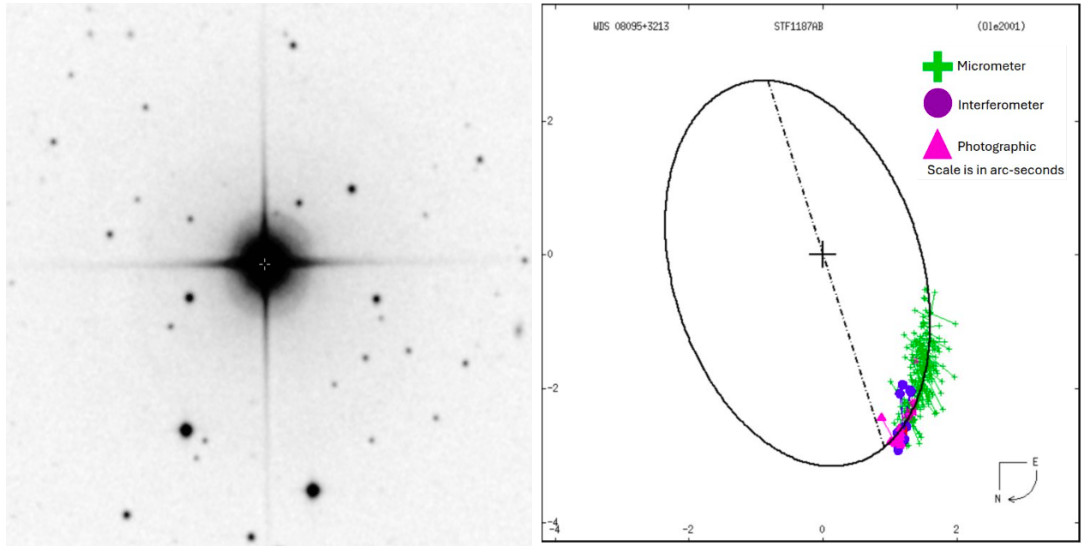

*Figure 1: Sloan Digital Sky Survey (SDSS) image of 08095+3213 (left) and its 6th Orbit plot (right).*

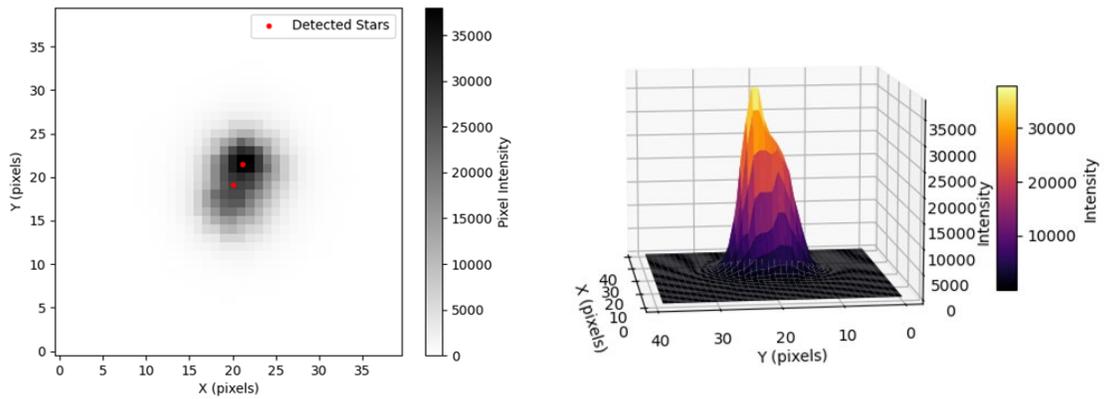

*Figure 2: Image of 08095+3213, observed using the JC-I filter (left) and the surface intensity plot created using PMDSA (right).*

*Table 5: Measurements of 08153-6255 using the i' filter.*

|  | ρ/" AIJ | ρ/" PMDSA | θ/° AIJ | θ/° PMDSA | ΔMag PMDSA |
|---|---|---|---|---|---|
| MEDIAN | 3.99 | 3.84 | 69.66 | 70.48 | 1.99 |
| MEAN | 4.00 | 3.81 | 69.53 | 70.32 | 2.00 |
| STDEV | 0.11 | 0.11 | 1.72 | 0.73 | 0.08 |
| SEM | 0.02 | 0.02 | 0.25 | 0.11 | 0.01 |

*Table 6: Measurements of 08153-6255 using the JC-V filter.*

|  | ρ/" AIJ | ρ/" PMDSA | θ/° AIJ | θ/° PMDSA | ΔMag PMDSA |
|---|---|---|---|---|---|
| MEDIAN | 3.96 | 3.76 | 69.73 | 69.79 | 2.08 |
| MEAN | 3.93 | 3.54 | 69.84 | 67.97 | 1.96 |
| STDEV | 0.15 | 0.76 | 1.86 | 12.64 | 0.44 |
| SEM | 0.02 | 0.12 | 0.26 | 2.05 | 0.07 |

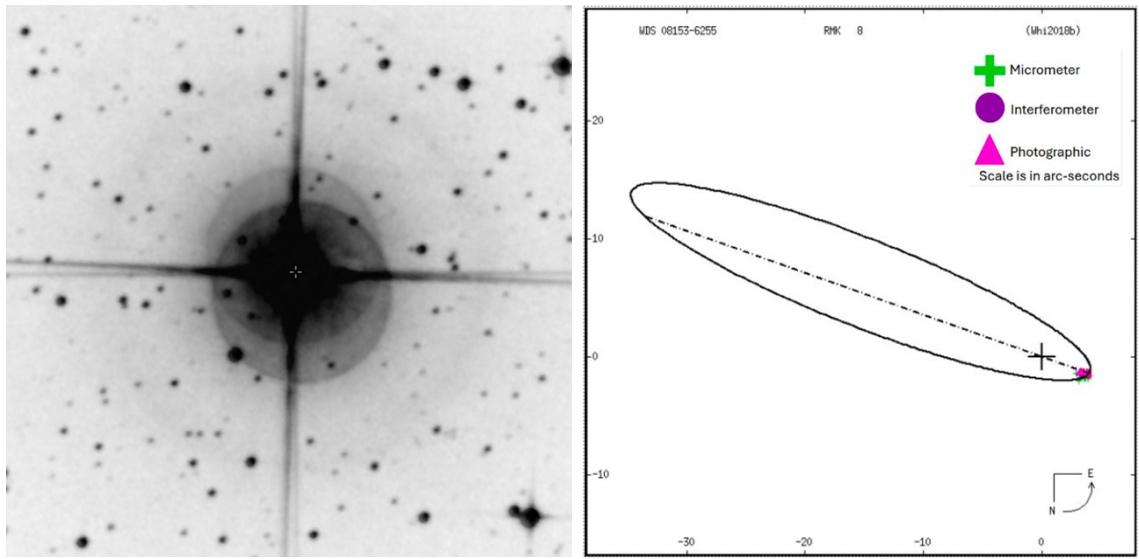

*Figure 3: SDSS image of 08153-6255 (left) and its 6th Orbit plot (right).*

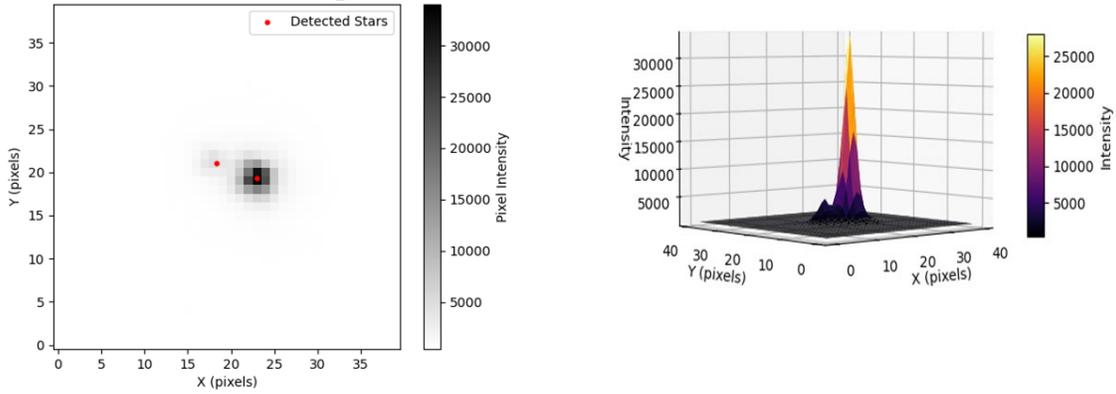

*Figure 4: Image of 08153-6255, observed using the i' filter (left) and the surface intensity plot created using PMDSA (right)*

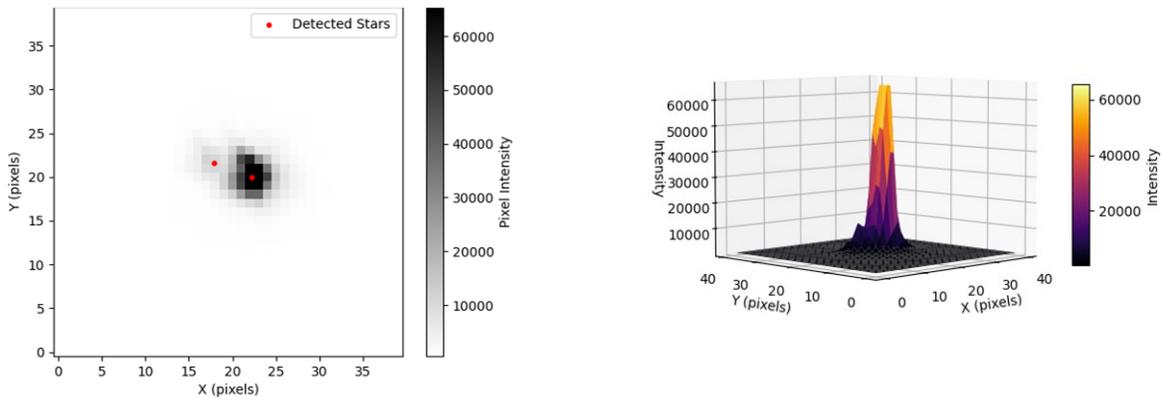

*Figure 5: Image of 08153-6255, observed using the JC-V filter (left) and the surface intensity plot created using PMDSA (right)*

*Table 7: Measurements of 08554+7048 using the JC-I filter.*

|  | ρ/" AIJ | ρ/" PMDSA | θ/° AIJ | θ/° PMDSA | ΔMag AIJ | Δmag PMDSA |
|---|---|---|---|---|---|---|
| MEDIAN | 3.45 | 3.63 | 359.56 | 359.38 | 0.10 | 0.10 |
| MEAN | 3.39 | 3.54 | 359.70 | 345.06 | 0.10 | 0.10 |
| STDEV | 0.26 | 0.29 | 0.81 | 49.57 | 0.06 | 0.06 |
| SEM | 0.05 | 0.06 | 0.16 | 9.91 | 0.01 | 0.01 |

*Table 8: Measurements of 08554+7048 using the JC-V filter.*

|  | ρ/" AIJ | ρ/" PMDSA | θ/° AIJ | θ/° PMDSA | ΔMag AIJ | Δmag PMDSA |
|---|---|---|---|---|---|---|
| MEDIAN | 3.40 | 3.42 | 361.50 | 361.45 | 0.15 | 0.15 |
| MEAN | 3.26 | 3.39 | 360.39 | 358.73 | 0.15 | 0.15 |
| STDEV | 0.36 | 0.32 | 1.09 | 1.47 | 0.03 | 0.03 |
| SEM | 0.07 | 0.07 | 0.22 | 0.30 | 0.01 | 0.01 |

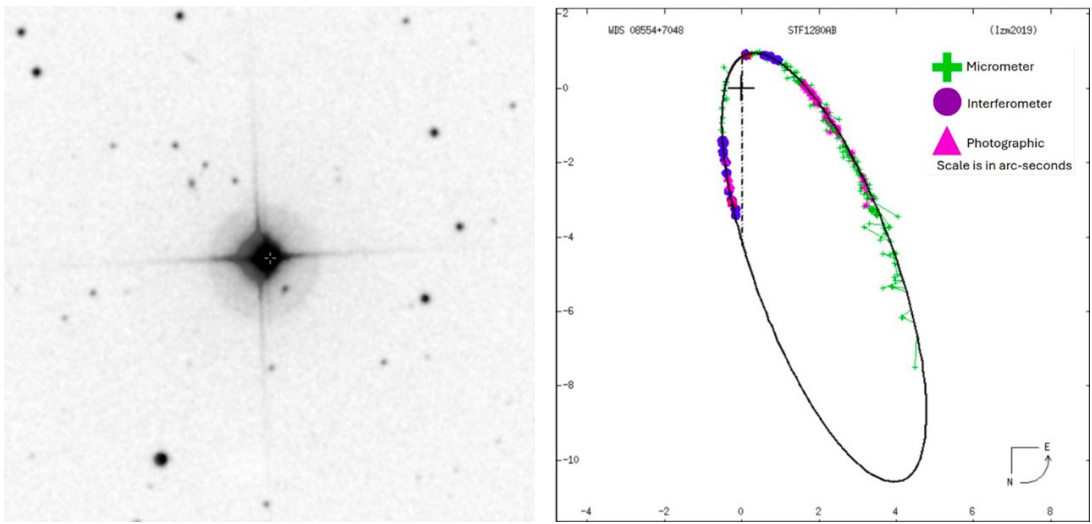

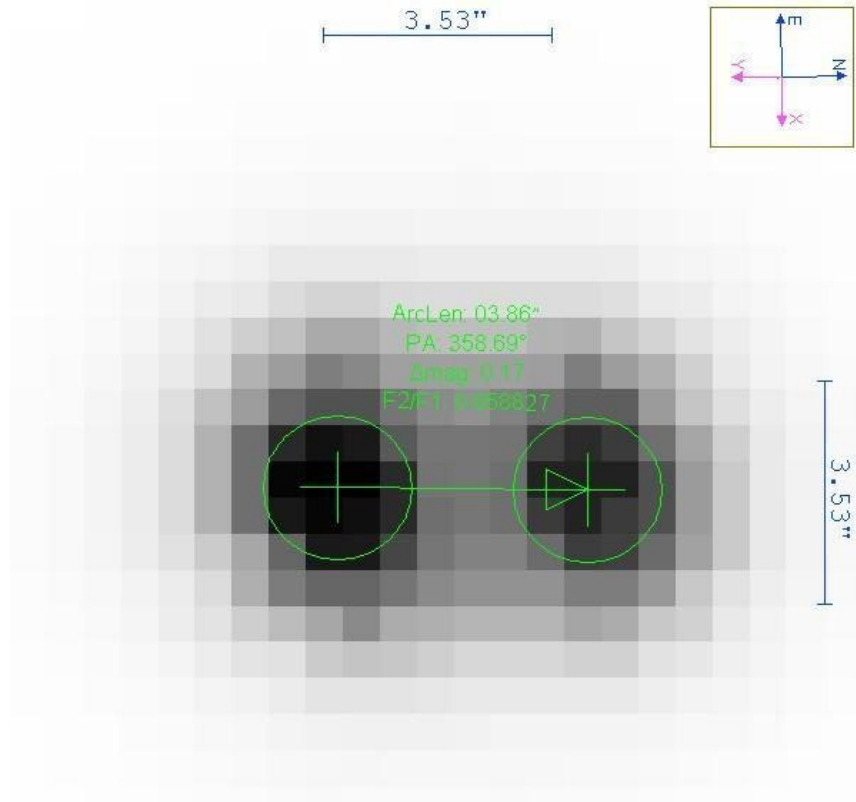

*Figure 7: Image of 08554+7048, successfully measured using AIJ.*

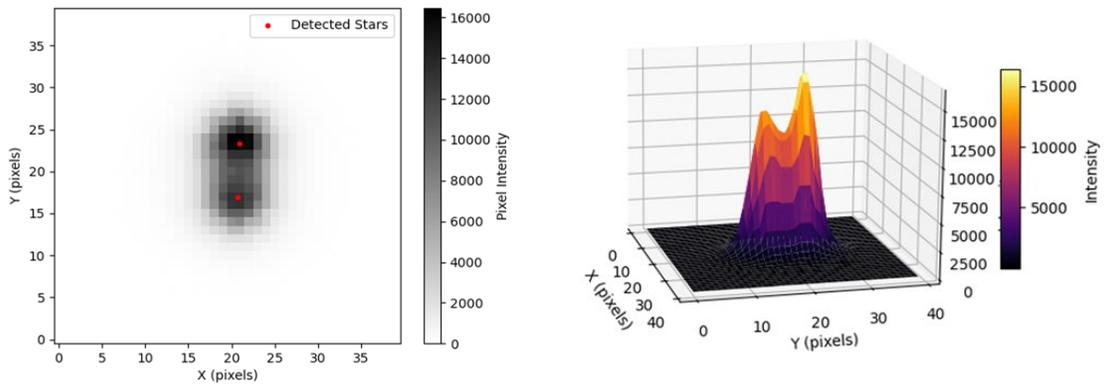

*Figure 8: Image of 08554+7048, observed using the JC-V filter (left) and the surface intensity plot created using PMDSA (right)*

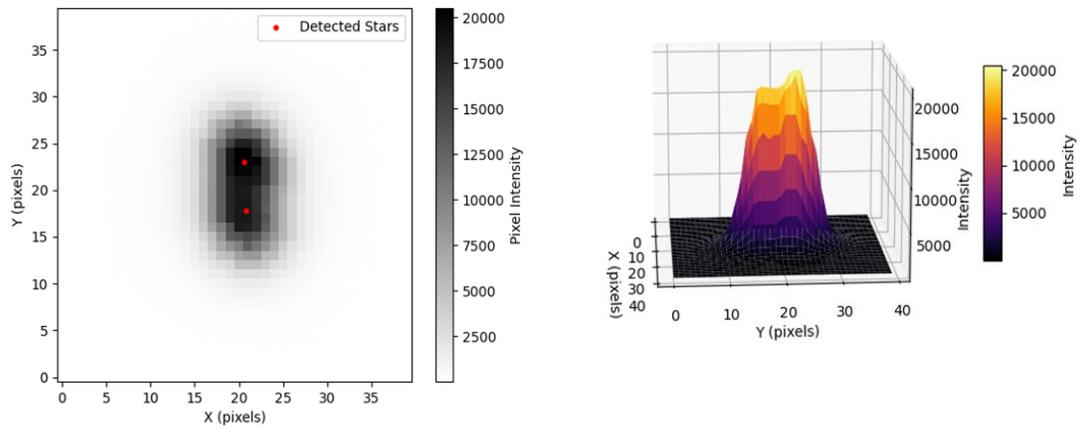

*Figure 9: Image of 08554+7048, observed using the JC-I filter (left) and the surface intensity plot created using PMDSA (right).*

*Table 9: A summary of median ρ, PA, and ΔMag values for all targets.*

| WDS ID | ρ/″ AIJ | ρ/″ PMDSA | θ/° AIJ | θ/° PMDSA | ΔMag AIJ | Δmag PMDSA |
|---|---|---|---|---|---|---|
| 08095+3213 | 2.10 | 1.94 | 19.13 | 21.76 | - | 0.41 |
| 08508+3504 | - | - | - | - | - | - |
| 08153-6255-i′ | 3.99 | 3.84 | 69.66 | 70.48 | - | 1.99 |
| 08153-6255-V | 3.96 | 3.76 | 69.73 | 69.79 | - | 2.08 |
| 10067+1754 | - | - | - | - | - | - |
| 08554+7048-I | 3.45 | 3.63 | 359.56 | 359.38 | 0.10 | 0.12 |
| 08554+7048-V | 3.40 | 3.42 | 361.50 | 361.45 | 0.15 | 0.19 |

## 5. Discussion

The two astrometric methods demonstrated commendable efficacy for all targets when measuring θ and ρ values. However, AstroImageJ was unable to offer ΔMag measurements in many cases because the images exhibited a blended point spread function. This meant that AIJ was unable to directly measure the difference in magnitude for several targets, and those for which it did, exhibited substantial deviation from expected values.

PMDSA used Python and the AstroPy module to analyze the data. It works by isolating the brightest sources closest to the center and using the coordinate points of those stars, along with information from the header of the images, to determine ρ, θ, and ΔMag. The code relies on World Coordinate System (WCS) information being included in the image metadata (Greisen, 2002). This allows it to determine the position of the stars, from which ρ and θ are measured. PMDSA employs the star detection algorithm developed by the Photutils detection module which utilizes the DAOStarfinder algorithm (Stetson, 1986). The code efficiently automates the measurement process and renders properly zoomed 2-dimensional images and 3-dimensional surface plots of the stars. Nevertheless, on multiple occasions, the code incorrectly identified the target stars (due to slight saturations or another star with a greater magnitude overshadowing the target stars), resulting in significant outliers within the dataset. These anomalies could easily be removed via manual pruning of the output, or possibly even statistical analysis.

The analysis confirmed that the instruments and methods used in this study are viable to measure separations between ~2.0" and ~15", given apertures, field of view (FOV), and camera capabilities. It is possible that an interferometric method using the same instruments and image trains would permit accurate measurement of this target and systems with Separations below 1.0". Unfortunately, WDS 10067+1754 was not measurable by either method, likely due to its very close separation (0.435"). Similarly, WDS 08508+3504 presented challenges that hindered our capacity to derive meaningful data. We postulate that the ΔMag was too small at the given separation of the system for our methods to isolate each star, even with visual methods.

For WDS 08095+3213, there is a discrepancy in the parallax of the system found in the Gaia database. The primary star (DR3 901953203859813888) has a listed parallax of 18.3667±0.025 milli-arc-seconds (mas) and a proper motion of 29.86 mas/year, while the secondary star (DR3 901953199567368960) has a listed parallax of 14.0074±0.900 mas with a proper motion of 30.36 mas/year. These values result in a physical separation of 16.94 parsecs (55.27 lightyears) and suggest a proper motion pair rather than a physical relation. We await Gaia's Data Release 4 to determine whether updated parallax measurements support our suggestion of a proper motion pair.

## 6. Conclusions

Compared to the extrapolated 6th Orbit catalog estimates, measurements using AstroImageJ were within 1.52% of θ and 14.87% of ρ, while PMDSA provided measurements within 4.09% of θ and 16.59% of ρ. Overall, AstroImageJ offered more accurate values for θ and ρ when compared to PMDSA. However, the PMDSA offered ΔMag values for all observed targets with an average error of 33.95% as opposed to AIJ with 56.90% for the targets it processed. Additionally, PMDSA offered faster processing speeds, processing approximately 50 plate-solved images in an average span of 3.5 minutes. Further comparisons of these astrometric methods are needed to identify which targets are measured most reliably and efficiently by each method. Additionally, many improvements can likely be made to PMDSA to enhance its accuracy.

**Acknowledgements**

This material is based upon work supported by the National Science Foundation under Grant No. 2428684 as part of the Four Corners Research Seminar organized by Paul McCudden.

Thanks to Rachel Freed, Ph.D. and Institute for Student Astronomical Research (InStAR) for guidance and instruction: https://www.in4star.org/

Thanks to Thomas C. Smith for guidance and instruction (Dark Ridge Observatory, Sacramento Mountains, New Mexico).

Thanks to Geoff Stone for guidance, instruction, and access to instruments (Dimension Point Observatory, Mayhill, NM).

Thanks to Rachel A. Matson, Ph.D., for access to the United States Naval Observatory historical data on target stars.

This work has made use of data from the European Space Agency (ESA) mission Gaia (https://www.cosmos.esa.int/gaia), processed by the Gaia Data Processing and Analysis Consortium (DPAC, https://www.cosmos.esa.int/web/gaia/dpac/consortium). Funding for the DPAC has been provided by national institutions, in particular the institutions participating in the Gaia Multilateral Agreement.

This work made use of Astropy: a community-developed core Python package and an ecosystem of tools and resources for astronomy (Astropy Collaboration, 2013, 2018, 2022).

This work made use of AstroImageJ: a graphical user interface (GUI) driven, public domain, Java-based, software package (Collins et al. 2017). https://www.astro.louisville.edu/software/astroimagej/

This research has made use of the Washington Double Star Catalog and the Sixth Catalog of Orbits of Visual Binary Stars maintained at the U.S. Naval Observatory.

This work made use of the SIMBAD service operated by Centre des Donnees Stellaires (Strasbourg, France), bibliographic references from the Astrophysics Data System maintained by SAO/NASA. https://simbad.u-strasbg.fr/simbad/

This research made use of the Stelle Doppie database, maintained by Gianluca Sordiglioni. https://www.stelledoppie.it/

This work made use of observations from the Las Cumbres Observatory global telescope (LCOGT) network. https://lco.global/

**Appendices**

PMDSA as used in this paper:
https://github.com/HarbingerOfFire/PMDSA/tree/a02e96ec50aa91667646e85265dd2c1ddeffcc24

*Table 10: Measurements for each image of 08095+3213 using the JC-V filter.*

| Index | ρ/" AIJ | ρ/" PMDSA | θ/° AIJ | θ/° PMDSA | ΔMag AIJ | Δmag PMDSA |
|---|---|---|---|---|---|---|
| 1 | 1.82 | 1.5 | 10.33 | 24.23 | - | 0.26 |
| 2 | 1.84 | 1.27 | 19.23 | 31.85 | - | 0.31 |
| 3 | 2.23 | 1.82 | 20..01 | 21.41 | - | 0.35 |
| 4 | 1.96 | - | 16.16 | - | - | - |
| 5 | 1.99 | 2.93 | 18.05 | 16.04 | - | 0.48 |
| 6 | 2.1 | 1.04 | 15.99 | 35 | - | 0.2 |
| 7 | 2.23 | 2.09 | 19.87 | 9.23 | - | 0.48 |
| 8 | 2.15 | 1.72 | 16.77 | 27.65 | - | 0.36 |
| 9 | 2.41 | 2.2 | 19.06 | 24.03 | - | 0.54 |
| 10 | 2.42 | 2.22 | 18.3 | 21.17 | - | 0.51 |
| 11 | 2.03 | 1.04 | 17.87 | 31.37 | - | 0.25 |
| 12 | 2.25 | 2.16 | 20.04 | 24.62 | - | 0.5 |
| 13 | 2.56 | 2.25 | 18.19 | 21.52 | - | 0.44 |
| 14 | 2.38 | 2.07 | 21.51 | 21.12 | - | 0.4 |
| 15 | 2.14 | 1.79 | 20.57 | 15.11 | - | 0.51 |
| 16 | 1.98 | 1.58 | 19.09 | 22.21 | - | 0.2 |
| 17 | 1.92 | 1.67 | 23.24 | 18.77 | - | 0.33 |
| 18 | 2.28 | 1.3 | 15.56 | 28.75 | - | 0.13 |
| 19 | 2.27 | 1.84 | 16.4 | 25.1 | - | 0.38 |
| 20 | 2.17 | 2.09 | 20.79 | 12.65 | - | 0.46 |
| 21 | 2.13 | 2.31 | 23.83 | 21.79 | - | 0.49 |

| Index | ρ/" AIJ | ρ/" PMDSA | θ/° AIJ | θ/° PMDSA | ΔMag AIJ | Δmag PMDSA |
|---|---|---|---|---|---|---|
| 22 | 2.05 | - | 22.07 | - | - | - |
| 23 | 2.13 | 2.03 | 21.52 | 25.4 | - | 0.33 |
| 24 | 2.11 | 3.08 | 18.77 | 11.45 | - | 0.52 |
| 25 | 2.15 | 1.59 | 19.43 | 29.1 | - | 0.33 |
| 26 | 1.78 | 1.33 | 17.62 | 17.56 | - | 0.42 |
| 27 | 2.03 | 1.75 | 19.13 | 21.72 | - | 0.5 |
| 28 | 2.31 | 2.5 | 21.07 | 22.3 | - | 0.46 |
| 29 | 2.27 | 1.52 | 16.41 | 32.69 | - | 0.43 |
| 30 | 2.35 | 2.16 | 15.29 | 24.96 | - | 0.54 |
| 31 | 1.91 | 2.48 | 19.3 | 11.65 | - | 0.3 |
| 32 | 2.18 | 3.36 | 14.41 | 15.2 | - | 0.82 |
| 33 | 2.06 | - | 16.14 | - | - | - |
| 34 | 1.81 | 0.85 | 23.31 | 28.24 | - | 0.29 |
| 35 | 2.1 | 1.35 | 17.65 | 28.41 | - | 0.21 |
| 36 | 1.73 | - | 17.2 | - | - | - |
| 37 | 2.06 | 1.7 | 20.54 | 28.18 | - | 0.32 |
| 38 | 1.42 | 2.83 | 22 | 11.62 | - | 0.5 |
| 39 | 2.24 | 1.82 | 17.89 | 17.39 | - | 0.56 |
| 40 | 2.17 | 2.43 | 17.54 | 17.65 | - | 0.32 |
| 41 | 2.16 | 1.56 | 21.53 | 24.38 | - | 0.44 |
| 42 | 1.84 | 2.05 | 24.84 | 18.09 | - | 0.21 |
| 43 | 1.69 | - | 14.33 | - | - | - |
| 44 | 1.71 | - | 27.85 | - | - | - |
| 45 | 2.34 | 3.09 | 16.17 | 7.33 | - | 0.42 |
| 46 | 1.88 | 1.17 | 26.78 | 26.2 | - | 0.38 |
| 47 | 1.78 | - | 24.36 | - | - | - |
| 48 | 1.67 | 2.89 | 25.92 | 11.75 | - | 0.5 |
| 49 | 1.71 | 2.35 | 25.2 | 12.01 | - | 0.26 |
| 50 | 1.44 | - | 21.86 | - | - | - |

*Table 11: Measurements for each image of 08554+7048 using the JC-I filter.*

| Index | ρ/" AIJ | ρ/" PMDSA | θ/° AIJ | θ/° PMDSA | ΔMag AIJ | Δmag PMDSA |
|---|---|---|---|---|---|---|
| 1 | 3.03 | 2.95 | 358.02 | 356.61 | - | 0.14 |
| 2 | 3.45 | 4.14 | 359.1 | 359.22 | 0.13 | 0.22 |
| 3 | 3.07 | 3.01 | 359.21 | 360.54 | 0.08 | 0.03 |
| 4 | 3.14 | 3.29 | 361.95 | 356.74 | - | 0.21 |
| 5 | 3.11 | 3 | 360.44 | 360.27 | - | 0.11 |
| 6 | 3.33 | 3.56 | 359.52 | 182.3 | 0.05 | 0.06 |
| 7 | 3.53 | 3.51 | 359.73 | 358.81 | 0.1 | 0.06 |
| 8 | 3.65 | 3.71 | 359.73 | 358.92 | 0.1 | 0.23 |
| 9 | 3.45 | 3.62 | 359.34 | 359.53 | 0.12 | 0.22 |
| 10 | 3.36 | 3.72 | 358.69 | 178.53 | 0.14 | 0.01 |
| 11 | 3.54 | 3.69 | 359.54 | 360.03 | 0.09 | 0.1 |
| 12 | 3.35 | 3.4 | 359.95 | 359.15 | 0.09 | 0.05 |
| 13 | 3.57 | 3.75 | 359.37 | 359.38 | 0.1 | 0.1 |
| 14 | 3.6 | 3.64 | 360.021 | 360.22 | 0.1 | 0.13 |
| 15 | 3.52 | 3.63 | 359.56 | 359.83 | 0.1 | 0.12 |
| 16 | 3.6 | 3.47 | 359.65 | 358.51 | 0.11 | 0.11 |
| 17 | 2.67 | 3.12 | 361.47 | 361.07 | 0.03 | 0.11 |
| 18 | 3.72 | 3.8 | 359.51 | 358.23 | 0.13 | 0.13 |
| 19 | 3.41 | 3.68 | 359.49 | 360.02 | 0.1 | 0.16 |
| 20 | 3.89 | 3.81 | 359.64 | 361.02 | 0.12 | 0.2 |
| 21 | 3.52 | 3.74 | 359.73 | 359.41 | 0.1 | 0.15 |
| 22 | 3.55 | 3.65 | 360.51 | 360.63 | 0.08 | 0.15 |
| 23 | 3.44 | 3.66 | 360.28 | 360.13 | 0.09 | 0.09 |
| 24 | 3.13 | 3.37 | 358.7 | 358.49 | 0.09 | 0.14 |
| 25 | 3.23 | 3.54 | 359.47 | 358.95 | 0.06 | 0.06 |

*Table 12: Measurements for each image of 08554+7048 using the JC-V filter.*

| Index | ρ/" AIJ | ρ/" PMDSA | θ/° AIJ | θ/° PMDSA | ΔMag AIJ | Δmag PMDSA |
|---|---|---|---|---|---|---|
| 1 | 3.4 | 3.66 | 1.5 | 1.45 | 0.132 | 0.21 |
| 2 | 3.67 | 3.64 | 0.39 | -1.27 | 0.18 | 0.27 |
| 3 | 3.58 | 3.66 | 0.68 | 2.12 | 0.17 | 0.23 |
| 4 | 2.76 | 3.56 | 2.9 | 2.6 | 0.11 | 0.23 |
| 5 | 2.93 | 3.52 | 1.04 | 0.3 | 0.13 | 0.25 |
| 6 | 3.47 | 3.65 | 1.16 | 0.87 | 0.15 | 0.22 |
| 7 | 3.51 | 3.76 | -1.12 | -1.03 | 0.17 | 0.24 |
| 8 | - | 3.92 | - | 0.03 | - | 0.19 |
| 9 | 3.06 | 3.49 | -0.98 | 1.73 | 0.11 | 0.13 |
| 10 | 3.52 | 3.12 | 1.48 | 1.28 | 0.15 | 0.11 |
| 11 | 2.6 | 3.37 | -0.41 | -0.66 | 0.09 | 0.15 |
| 12 | 3.52 | 2.75 | 0.63 | -0.96 | 0.18 | 0.23 |
| 13 | 2.93 | 3.63 | -0.06 | 1.76 | 0.12 | 0.18 |
| 14 | 3.52 | 3.29 | -0.52 | 1.47 | 0.17 | 0.17 |
| 15 | 3.52 | 3.32 | -0.51 | -0.74 | 0.17 | 0.14 |
| 16 | 3.25 | 3.29 | 0.28 | -1.1 | - | 0.23 |
| 17 | 3.65 | 3.27 | 0.05 | -2.82 | 0.17 | 0.18 |
| 18 | 2.8 | 2.65 | 2.07 | -1.2 | - | 0.07 |
| 19 | 3.39 | 3.25 | 1.63 | 2.88 | 0.13 | 0.12 |
| 20 | 2.82 | 3.4 | -0.88 | 0.36 | 0.08 | 0.04 |
| 21 | 3.57 | 3.44 | -1.65 | -1.43 | 0.19 | 0.21 |
| 22 | 3.69 | 3.75 | -0.13 | -0.41 | 0.17 | 0.2 |
| 23 | 3.1 | 2.86 | 0 | -0.35 | - | 0.16 |
| 24 | 3.37 | 3.21 | 0.69 | -0.49 | - | 0.13 |
| 25 | 2.54 | - | -0.14 | - | - | - |

*Table 13: Measurements for each image of 08153-6255 using the i' filter.*

| Index | ρ/" AIJ | ρ/" PMDSA | θ/° AIJ | θ/° PMDSA | ΔMag AIJ | Δmag PMDSA |
|---|---|---|---|---|---|---|
| 1 | 3.93 | 3.67 | 70.5 | 69.6 | - | 1.97 |
| 2 | 3.9 | 3.66 | 70.17 | 69.57 | - | 1.97 |
| 3 | 3.8 | 3.92 | 69.05 | 70.66 | - | 1.99 |
| 4 | 3.77 | 3.93 | 68.82 | 70.53 | - | 1.99 |
| 5 | 4.07 | 3.92 | 70.04 | 70.8 | - | 2.06 |
| 6 | 4.16 | 3.93 | 73.14 | 70.65 | - | 2.06 |
| 7 | 3.84 | - | 73.74 | - | - | - |
| 8 | 3.99 | - | 70.01 | - | - | - |
| 9 | 4.1 | 3.9 | 70.89 | 70.48 | - | 1.96 |
| 10 | 4.06 | 3.91 | 72.15 | 70.32 | - | 1.96 |
| 11 | 4.02 | 3.83 | 71.12 | 70.18 | - | 2.09 |
| 12 | 3.9 | 3.84 | 71.15 | 70.08 | - | 2.09 |
| 13 | 3.97 | 3.7 | 66.95 | 70 | - | 1.99 |
| 14 | 3.93 | 3.71 | 68.25 | 69.82 | - | 1.99 |
| 15 | 4.05 | 3.94 | 66.65 | 70.36 | - | 2.03 |
| 16 | 4.11 | 3.95 | 69.27 | 70.18 | - | 2.03 |
| 17 | 3.92 | 3.75 | 68.99 | 71.14 | - | 1.87 |
| 18 | 3.88 | 3.76 | 69.66 | 70.99 | - | 1.87 |
| 19 | 3.99 | 3.58 | 71.44 | 70.89 | - | 1.8 |
| 20 | 3.99 | 3.58 | 72.25 | 70.78 | - | 1.8 |
| 21 | 4.08 | 3.93 | 69.1 | 70.48 | - | 2.02 |
| 22 | 4 | 3.94 | 70.17 | 70.34 | - | 2.03 |
| 23 | 4.08 | - | 71.27 | - | - | - |
| 24 | 4.01 | - | 70.02 | - | - | - |
| 25 | 3.92 | 3.68 | 72.03 | 69.73 | - | 1.99 |
| 26 | 3.96 | 3.69 | 69.56 | 69.52 | - | 1.99 |
| 27 | 4.02 | 3.88 | 66.25 | 70.99 | - | 2.06 |
| 28 | 3.96 | 3.89 | 67.28 | 70.86 | - | 2.06 |

| Index | ρ/" AIJ | ρ/" PMDSA | θ/° AIJ | θ/° PMDSA | ΔMag AIJ | Δmag PMDSA |
|---|---|---|---|---|---|---|
| 29 | 3.9 | 3.78 | 67.62 | 69.1 | - | 2.1 |
| 30 | 4.03 | 3.79 | 68.24 | 68.96 | - | 2.1 |
| 31 | 3.94 | 3.59 | 67.93 | 68.41 | - | 2.06 |
| 32 | 3.88 | 3.61 | 66.2 | 68.33 | - | 2.06 |
| 33 | 4.05 | 3.76 | 68.02 | 70.48 | - | 1.9 |
| 34 | 4.03 | 3.78 | 68.32 | 70.42 | - | 1.9 |
| 35 | 4.09 | 3.82 | 70.09 | 70.84 | - | 1.99 |
| 36 | 4.23 | 3.83 | 70.01 | 70.69 | - | 2 |
| 37 | 3.88 | - | 66.39 | - | - | - |
| 38 | 3.96 | - | 67.37 | - | - | - |
| 39 | 4.2 | 3.9 | 70.39 | 69.57 | - | 2.01 |
| 40 | 3.96 | 3.92 | 69.59 | 69.48 | - | 2.01 |
| 41 | 4.19 | 3.83 | 70.53 | 70.77 | - | 2.09 |
| 42 | 4.11 | 3.84 | 70.02 | 70.6 | - | 2.09 |
| 43 | 4 | 3.85 | 69.07 | 70.99 | - | 1.97 |
| 44 | 3.96 | 3.86 | 70.1 | 70.85 | - | 1.97 |
| 45 | 4.01 | 3.77 | 69.94 | 71.13 | - | 1.95 |
| 46 | 4.16 | 3.92 | 68.77 | 70.65 | - | 2.16 |
| 47 | 4.28 | 3.93 | 68.97 | 70.55 | - | 2.17 |
| 48 | 3.98 | 3.84 | 69.5 | 71.45 | - | 1.91 |
| 49 | 3.99 | 3.85 | 70.02 | 71.33 | - | 1.91 |

*Table 14: Measurements for each image of 08153-6255 using the JC-V filter.*

| Index | ρ/" AIJ | ρ/" PMDSA | θ/° AIJ | θ/° PMDSA | ΔMag AIJ | Δmag PMDSA |
|---|---|---|---|---|---|---|
| 1 | 3.77 | 3.41 | 69.25 | 69.5 | - | 2.08 |
| 2 | 3.73 | 3.42 | 66.67 | 69.46 | - | 2.08 |
| 3 | 4 | 3.88 | 69.73 | 69.48 | - | 2.15 |
| 4 | 3.68 | 3.88 | 68.87 | 69.45 | - | 2.15 |
| 5 | 3.76 | 3.2 | 70.12 | 66.66 | - | 2.16 |
| 6 | 4.12 | 3.22 | 71.29 | 66.52 | - | 2.15 |
| 7 | 4 | 3.42 | 70.68 | 77.22 | - | 1.78 |
| 8 | 3.8 | 3.42 | 70.82 | 77.08 | - | 1.78 |
| 9 | 3.72 | - | 72.32 | - | - | - |
| 10 | 3.95 | - | 72.15 | - | - | - |
| 11 | 3.87 | 3.8 | 72.79 | 73.61 | - | 1.91 |
| 12 | 4.04 | 3.83 | 69.5 | 73.56 | - | 1.91 |
| 13 | 3.73 | 3.78 | 72.41 | 69.2 | - | 2.1 |
| 14 | 3.8 | 3.8 | 70.6 | 69.06 | - | 2.1 |
| 15 | 3.83 | - | 67.12 | - | - | - |
| 16 | 3.99 | - | 66.65 | - | - | - |
| 17 | 3.92 | 3.68 | 69.85 | 68.44 | - | 2.06 |
| 18 | 4.01 | 3.7 | 69.15 | 68.31 | - | 2.06 |
| 19 | 3.96 | 3.84 | 69.87 | 69.64 | - | 1.94 |
| 20 | 4.02 | 3.86 | 68.07 | 69.57 | - | 1.94 |
| 21 | 3.91 | 3.71 | 70.98 | 74.73 | - | 1.97 |
| 22 | 3.98 | 3.81 | 71.25 | 69.75 | - | 2.08 |
| 23 | 3.99 | - | 72.27 | - | - | - |
| 24 | 3.95 | - | 72.14 | - | - | - |
| 25 | 3.96 | 3.76 | 68.72 | 70.92 | - | 2.11 |
| 26 | 4.03 | 3.77 | 68.34 | 70.83 | - | 2.11 |
| 27 | 3.98 | 3.53 | 69.92 | 73.34 | - | 2.11 |
| 28 | 3.97 | 3.54 | 73.74 | 73.24 | - | 2.11 |

| Index | ρ/" AIJ | ρ/" PMDSA | θ/° AIJ | θ/° PMDSA | ΔMag AIJ | Δmag PMDSA |
|---|---|---|---|---|---|---|
| 29 | 3.92 | 3.75 | 69.73 | 75.03 | - | 2 |
| 30 | 3.91 | 3.76 | 68.59 | 74.83 | - | 2 |
| 31 | 3.9 | 3.56 | 69.7 | 68.66 | - | 1.95 |
| 32 | 3.97 | 3.73 | 67.61 | 71.21 | - | 2.01 |
| 33 | 4.08 | - | 69.09 | - | - | - |
| 34 | 4.08 | - | 68.87 | - | - | - |
| 35 | 3.93 | - | 70.84 | - | - | - |
| 36 | 3.96 | - | 69.33 | - | - | - |
| 37 | 4.06 | 3.86 | 66.09 | 69.47 | - | 2.17 |
| 38 | 3.85 | 3.88 | 66.67 | 69.34 | - | 2.17 |
| 39 | 4.19 | 3.64 | 70.69 | 69.82 | - | 2.15 |
| 40 | 4.21 | 3.66 | 71.89 | 69.71 | - | 2.15 |
| 41 | 4.28 | - | 69.73 | - | - | - |
| 42 | 3.92 | - | 71.61 | - | - | - |
| 43 | 3.98 | 4.01 | 71.58 | 71.25 | - | 2.08 |
| 44 | 4.12 | 4.03 | 71.48 | 71.13 | - | 2.08 |
| 45 | 4.02 | 3.87 | 68.26 | 70.25 | - | 2.13 |
| 46 | 3.6 | 3.89 | 68.68 | 70.11 | - | 2.13 |
| 47 | 3.52 | 3.79 | 66.9 | 70.16 | - | 2.19 |
| 48 | 3.93 | 3.81 | 72.76 | 70.08 | - | 2.19 |
| 49 | 3.94 | 0.46 | 68.64 | 16.61 | - | 0.16 |
| 50 | 3.83 | 0.46 | 68.01 | 15.58 | - | 0.16 |

*Table 15: Gaia Parameters of the targets.*

| Star Name | 08095+3213 Pri | 08095+3213 Sec | 08554+7048 Pri | 08554+7048 Sec | 08153-6255 Pri | 08153-6255 Sec |
|---|---|---|---|---|---|---|
| source_id | 901953203859813888 | 901953199567368960 | 1118916397395366656 | 1118916397393675648 | 5277370356917351808 | 5277370356917351936 |
| ra | 122.38 | 122.38 | 133.84 | 133.84 | 123.82 | 123.82 |
| dec | 32.22 | 32.22 | 70.79 | 70.79 | -62.92 | -62.92 |
| pm | 30.36 | 29.86 | 1364.13 | 1417.05 | 26.61 | 25.37 |
| pmra | 30.16 | 29.85 | -1329.02 | -1353.98 | -23.54 | -20.96 |
| pmdec | -3.49 | -0.59 | -307.49 | -418.03 | -12.41 | -14.29 |
| parallax | 18.37 | 14.01 | 86.35 | 86.30 | 12.72 | 12.83 |
| dist | 54.45 | 71.39 | 11.58 | 11.59 | 78.63 | 77.95 |
| phot_g_mean_mag | 7.05 | 7.79 | 8.27 | 8.07 | 5.24 | 7.65 |
| phot_bp_mean_mag | 7.23 | 7.99 | 9.13 | 8.89 | 5.27 | 7.82 |
| phot_rp_mean_mag | 6.65 | 7.40 | 7.34 | 7.17 | 5.16 | 7.21 |
| bp_rp | 0.59 | 0.59 | 1.79 | 1.72 | 0.11 | 0.61 |
| radial_velocity | | | | | -8.38 | |
| teff_gspphot | 6483.16 | 6420.51 | 3711.76 | 3950.40 | 8600.62 | 6720.19 |
| distance_gspphot | 54.89 | 71.23 | 11.59 | 11.57 | 83.94 | 77.99 |
| radius_gspphot | 1.48 | 1.38 | 0.69 | 0.65 | 3.10 | 1.59 |